# Ripple morphology of graphitic surfaces: a comparison between few-layer graphene and HOPG

N. Haghighian[a], D. Convertino[b], V. Miseikis[b], F Bisio[c], A. Morgante[d,e], C. Coletti[b,f], M. Canepa[a], O. Cavalleri[g]

The surface structure of Few-Layer Graphene (FLG) epitaxially grown on the C-face of SiC has been investigated by TM-AFM in ambient air and upon interaction with diluted aqueous solutions of bio-organic molecules (dimethyl sulfoxide, DMSO, and L-Methionine). On pristine FLG we observe nicely ordered, three-fold oriented rippled domains, with a 4.7±0.2 nm periodicity (small periodicity, SP) and a peak-to-valley distance in the range 0.1÷0.2 nm. Upon mild interaction of the FLG surface with the molecular solution, the ripple periodicity "relaxes" to 6.2±0.2 nm (large periodicity, LP), while the peak-to-valley height increases to 0.2÷0.3 nm. When additional energy is transferred to the system through sonication in solution, graphene planes are peeled off from FLG, as shown by quantitative analysis of XPS and Raman spectroscopy data which indicate a neat reduction of thickness. Upon sonication rippled domains are no longer observed. Regarding HOPG, we could not observe ripples on cleaved samples in ambient air, while LP ripples develop upon interaction with the molecular solutions. Recent literature on similar systems is not univocal regarding the interpretation of rippling. The complex of our comparative observations on FLG and HOPG can be hardly rationalized solely on the base of surface assembly of molecules, either organic molecules coming from the solution or adventitious species. We propose to consider the ripples as the manifestation of the free-energy minimization of quasi-2D layers, eventually affected by factors such as the interplane stacking, the interaction with molecules and/or with the AFM tip.

[a.] OptMatLab, Dipartimento di Fisica, Università di Genova, via Dodecaneso 33, 16146 Genova, Italy
[b.] CNI@NEST, Istituto Italiano di Tecnologia, Piazza S. Silvestro 12, 56127 Pisa, Italy
[c.] CNR-SPIN, C.so Perrone 24, 16152 Genova, Italy
[d.] CNR-IOM, Strada Statale 14 – km 163.5, 34149 Trieste, Italy
[e.] Dipartimento di Fisica, Università di Trieste, Via Valerio 2, 34127 Trieste, Italy
[f.] Graphene Labs, Istituto Italiano di Tecnologia, via Morego 30, 16163 Genova, Italy
[g.] Dipartimento di Fisica, Università di Genova, via Dodecaneso 33, 16146 Genova, Italy. E-mail: cavalleri@fisica.unige.it

## Introduction

Interfacial phenomena occurring at the surface of graphite or 2D graphitic compounds interacting with organic solvent/solutions present many aspects of interest in diverse fields. Relevant examples regard the development of bio-sensing devices[1-11] and the liquid-phase exfoliation of graphite[12-20] which is of prominent importance for cost-effective, large scale exploitation of graphene[21-25].

We report here on a subtle and interesting issue related to the interaction of aqueous solutions of organic and biologic molecules with graphitic surfaces.

In an early report[26], we investigated by AFM the surface morphology resulting from the interaction of diluted solutions of several proteins with the surface of cleaved HOPG. Regardless of the protein structure, interaction with the solution led to the observation of nicely ordered "ripples", showing a well-defined periodicity of 6.2 nm. The "ripples", forming extended three-fold oriented nanopatterned domains, were tentatively ascribed to the re-assembly of peptides following the unfolding of proteins[26]. More recent experiments we performed on HOPG led to the observation of the same type of rippled domains upon interaction with several organic and biologic molecules, such as polyelectrolytes[27], ε-caprolactam[28] or even small molecules like L-methionine, much simpler than proteins[27]. These new evidences cast doubts on an interpretation of ripples based solely on the simple assembly of molecules.

Rippled domains with seemingly similar structure but with periodicity of about 4-5 nm have been reported in recent literature on a variety of graphitic surfaces. Hwang and coworkers observed ripples at the water/HOPG interface[29-31] and at the interface between water and graphene-coated mica[32] and assigned the ripples to ordered gas domains, formed after diffusion of water-dissolved gas molecules towards the hydrophobic surface. Similar ripples were observed on bilayer graphene[33] and on hydrogen-intercalated graphene[34] on SiC, exposed to ambient air. Assuming the presence of a wetting layer on the air–exposed surface, the authors endorsed the interpretation of Hwang and coworkers and assigned the ripples to adsorbed gas layers.

The interpretation in terms of airborne molecular adsorption has been also accepted in a most recent report dealing with friction properties of 2D materials[35] where ripples were observed on exfoliated graphene and exfoliated hBN deposited on $SiO_2$ as well as on epitaxial graphene/hBN heterostructures exposed to ambient air. Rippled domains were identified as the cause of friction anisotropy in other AFM studies on exfoliated graphene[36-39] and on other 2D materials, like $MoS_2$, $NbSe_2$ and hBN, on weakly adherent substrates[38]. Friction anisotropy was found to decrease when the applied load was increased[37] whereas an increase in friction was observed when the sheet thickness decreased to one monolayer[38]. These reports assigned friction anisotropy to rippled domains resulting from out-of-plane deformations of ultrathin and weakly interacting films, i.e. almost 2D systems.

The astonishingly similar ripple morphology which is observed in experiments that are apparently very different renewed our interest in the rippling of graphitic surfaces. We opted to perform new experiments looking at the interaction of organic molecules with so-called epitaxial, few-layer graphene, grown on the carbon-rich surface of SiC (in brief FLG). Owing to the very weak inter-plane electronic coupling, FLG can be considered as an ultrathin stack of "independent" graphene planes[40]. Thanks to the lack of bulk graphitic signal, FLG allows easier detection of eventual removal of graphene layers after interaction with molecules, which can be easily observed as a film thickness reduction, through methods like XPS and Raman analysis. In this respect, FLG, better than HOPG, allows to explore the correlation between ripple formation and possible exfoliation of graphene planes. In addition, the comparison of results obtained on FLG and on HOPG can shed light on the role of stacking of graphene planes in the rippling process. In facts, the relationship between molecular adsorption and plane stacking deserves attention when considering interfacial phenomena on graphitic compounds as emphasized e.g. by a recent paper on trilayer graphene on $SiO_2$ which reported a transformation of ABC-stacked to ABA-stacked domains upon deposition in vacuum of triazine molecules[41].

In this work we shed further light on the formation and nature of the ripples on HOPG and graphene focussing on the origin of both the small periodicity (SP, 4.7 nm) and the larger periodicity (LP, 6.2 nm) ripple morphology. To this end we performed experiments with aqueous solutions of dimethyl sulfoxide (DMSO), a solvent widely exploited for liquid phase exfoliation of graphene[42] and with solutions of the aminoacid L-methionine, chosen as an example of a simple biomolecule with a non-polar side chain and two ionisable, polar groups.

The results obtained by simply dropping molecular solutions onto the FLG surface have been compared to those obtained after delicate exfoliation through sonicating the surface into a diluted DMSO solution. We will show that "gentle" interaction with dropped DMSO or L-methionine produces a drastic change of the rippling morphology of FLG which relaxes from a SP to a LP structure, i.e. the same obtained in experiments on HOPG.

Instead, exfoliation, i.e. a harder treatment, destroys the surface order and the ripple morphology.

## Experimental

### Materials

FLG was grown on the C-terminated face of insulating on-axis-oriented silicon carbide 4H-SiC(0001) (0.5 × 0.5 cm$^2$) substrates in a resistively heated cold-wall reactor (Aixtron HT-BM) via thermal decomposition[43,44]. Before growth, the samples were hydrogen etched in an $H_2$/Ar gas flow (500/500 sccm), in order to obtain atomically flat surfaces[45]. The pressure was 450 mbar, temperature and etch time were 1250 °C and 5 min. Growth was performed in the same reactor in an argon atmosphere of 780 mbar at a temperature and growth time of 1350 °C and 15 min. The quality of the samples and the number of graphene layers were assessed by Raman spectroscopy (as detailed below in the paper). Highly oriented pyrolytic graphite, HOPG, (12x12x1.7 mm$^3$, Grade ZYB) was purchased from NT-MDT, Russia.

Solutions were prepared by dissolving DMSO, (Sigma-Aldrich, 99%) and L-Methionine (Sigma-Aldrich, 98%) in Milli-Q water to a final concentration of 0.3 µg/ml. Both compounds were used as received without further purification. The interaction of the two kinds of solution with the carbon surface of FLG and freshly cleaved HOPG was achieved through two methods:

(i) "Mild" treatment (dropping)

DMSO or L-Methionine solutions were dropped on the surface. After one hour incubation at room temperature, samples were thoroughly rinsed with Milli-Q and dried under a nitrogen stream.

(ii) "Hard" treatment (sonication)

Glass vials containing molecular solutions (DMSO or L-methionine) and the substrate were placed in an ultrasonic bath. Samples were sonicated for 1 h. After sonication the samples were thoroughly rinsed with Milli-Q water and dried under a nitrogen stream.

In order to observe exfoliation products from graphite through UV-Vis absorption, freshly peeled HOPG flakes were sonicated in either DMSO or L-methionine solution for time intervals from 30 min to 2 hours.

### Characterization methods

The morphology of pristine and processed surfaces were investigated by atomic force microscopy (AFM). Tapping mode AFM measurements were performed using a Multimode/Nanoscope IV system (Bruker) and Si cantilevers (OMCL-AC160 TS, Olympus) with a nominal tip radius of 7 nm and a resonance frequency in air of about 330 kHz.

The change in FLG thickness induced by exfoliation was studied by X-ray photoelectron spectroscopy (XPS) and Raman spectroscopy.

XPS measurements were carried out using a 5600 MultiTechnique apparatus operated as reported in previous studies[46]. An X-ray Al-monochromatised source (hν=1486.6 eV) was used. The spectra are shown as a function of binding energy (BE): the scale was referenced to the C1s signal of C-C adventitious carbon on the SiC substrate set at 284.9 eV. The take-off angle was 45°.

Raman spectroscopy measurements were performed by using a Renishaw InVia system equipped with a 532 nm green laser and a motorized stage for large-area mapping. A beam spot size of approximately 1µm in diameter was used.

UV-Vis absorption spectroscopy measurements were performed using a Jasco V-530 double-beam spectrophotometer.

## Results

### Surface Morphology: SP and LP ripples

**Pristine FLG.** Figure 1a shows a representative, large-scale tapping mode AFM image of the surface of pristine

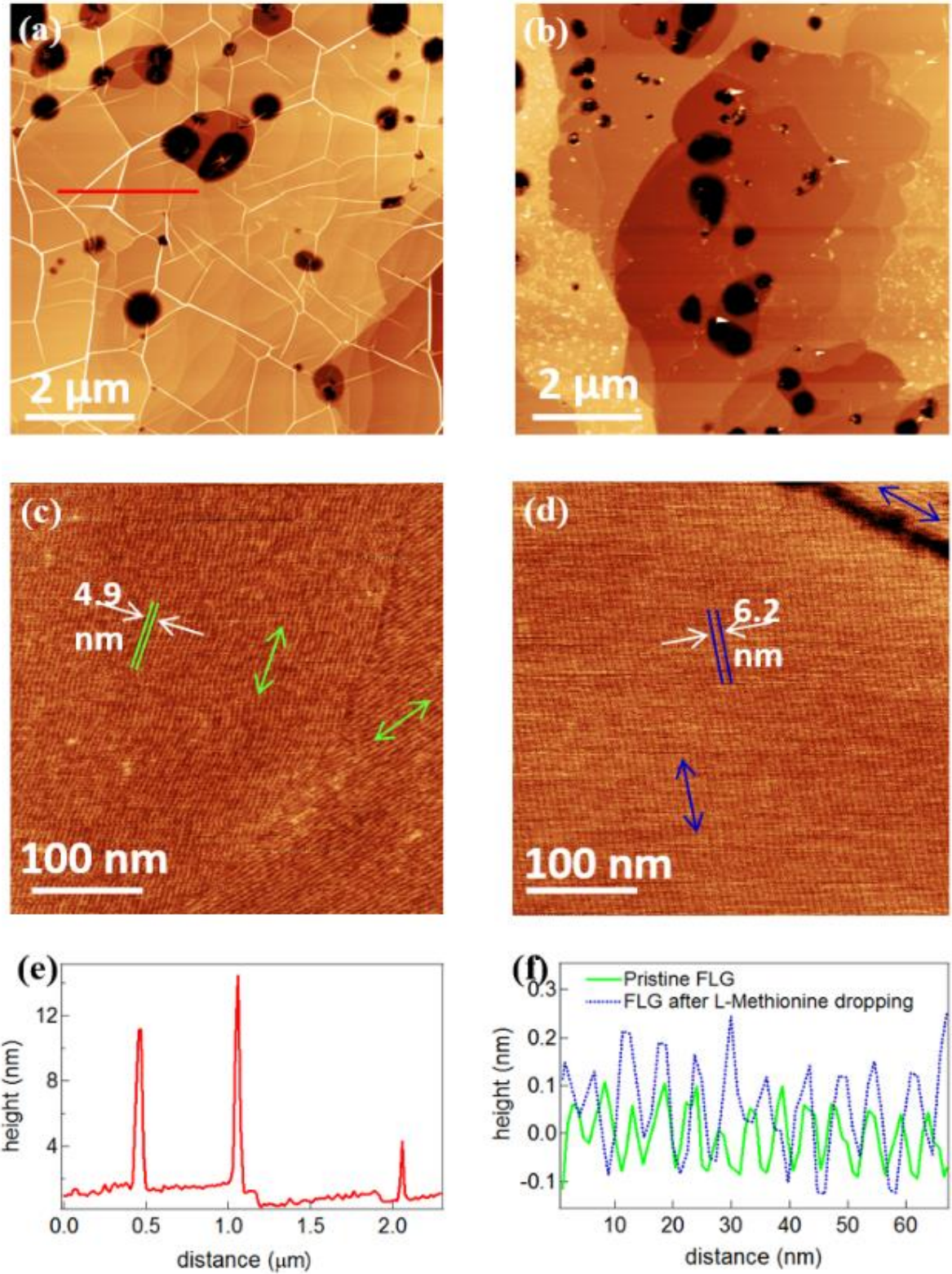

Figure 1 Tapping mode AFM height images of FLG on SiC. (a) Pristine FLG; Z-scale: 25 nm (b) FLG after sonication in DMSO solution; Z-scale: 25 nm (c) Pristine FLG: zooming in the regions delimited by ridges; Z-scale: 0.8 nm (d) FLG after dropping with L-Methionine solution; Z-scale: 0.8 nm. In panels (c) and (d) arrows and double segments indicate the directions and spacing of the ripples. (e) Z profile obtained along the red line of image in panel (a). (f) Z profile of ripples on pristine FLG (green line) and after methionine dropping (blue line).

FLG. The surface is characterized by micrometer-sized flat domains separated by thin ridges. Deep holes, 20-30 nm in depth, can be also observed, which likely result from the graphitization. As inferred from the analysis of the z profiles (Figure 1e), ridges have different height, from a few up to 10-15 nanometers. They are arranged according to a pseudo hexagonal network, suggesting their alignment along high-symmetry directions of the FLG film. The ridges run continuously over the underneath SiC steps suggesting that the graphene films are likely to be continuous over the micrometer distance. Ridges are typical of FLG on SiC[47-49]; a similar morphology has been observed also on FLG grown on Ni substrates by chemical vapour deposition[50].

According to literature, the ridges are due to folds of graphene layers which form during cooling[47,49]. The formation of ridges is due to the relaxation of the anisotropic compressive stress of the film resulting from the different thermal contraction of graphene and substrate.

AFM images of pristine FLG samples obtained by zooming in the flat regions delimited by ridges show the presence of well-defined, regular nanopatterned (rippled) domains (Figure 1c). Ripples are oriented along the high-symmetry directions of the carbon planes. Analysis of the z-profiles of the rippled domains (a representative example is shown in Figure 1f, green line) indicates a periodicity of 4.7±0.2 nm (SP ripples) with peak-to-valley height of 0.1÷0.2 nm. Ripples could not be detected on the graphene grown on the Si face of SiC. This finding could be related to the stronger interaction between graphene and substrate mediated by the presence of the buffer layer[51]. The observation of a ripple morphology on pristine FLG (carbon face) appears at sharp variance with previous observations on pristine (cleaved) HOPG. Indeed, rippled domains were observed on HOPG only upon interaction with a aqueous molecular solution[26].

**"Mild" treatment with DMSO/L-methionine.** We have executed AFM experiments after dropping either DMSO or L-methionine solutions on the FLG surface. No morphological changes were observed at the mesoscopic scale. Interestingly, a change of the ripple morphology was observed at the nanometer scale. For both DMSO and L-methionine the ripple periodicity "relaxes" from 4.7±0.2 nm to 6.2±0.2 nm, while the peak-to-valley height increases to 0.2÷0.3 nm (Figure 1f, dotted blue line). The SP to LP structural change was triggered by the interaction with the solution.

Regarding HOPG, as shown in Figure 2a on the example of DMSO, dropping of DMSO or L-methionine induced wide, three-fold oriented LP rippled domains with the same periodicity observed in previous experiments using other compounds; we note that in some experiments dealing with proteins both SP and LP ripples were observed[26].

**"Hard" treatment: sonication in DMSO.** Sonication of pristine FLG samples in the DMSO solution significantly affects the surface morphology. Large-scale AFM measurements shown in Figure 1b indicate the removal of ridges, that is a rather strong indication of a peeling process. Exfoliation could be also invoked as the origin of flakes or debris (small bright irregular "islands" in Figure 1b) that decorate many flat regions. Going to the nanoscale, we observed the complete removal of ripples (data not shown). Control AFM experiments on HOPG after sonication in DMSO (Figure 2b) showed irregular flakes, likely resulting from peeling and re-stacking processes. The height of the flakes varies from the equivalent of a few up to a few tens of carbon layers. No ripple structure was observed at the nanoscale.

No rippled domains could be therefore detected on both FLG and HOPG samples after sonication in DMSO solution.

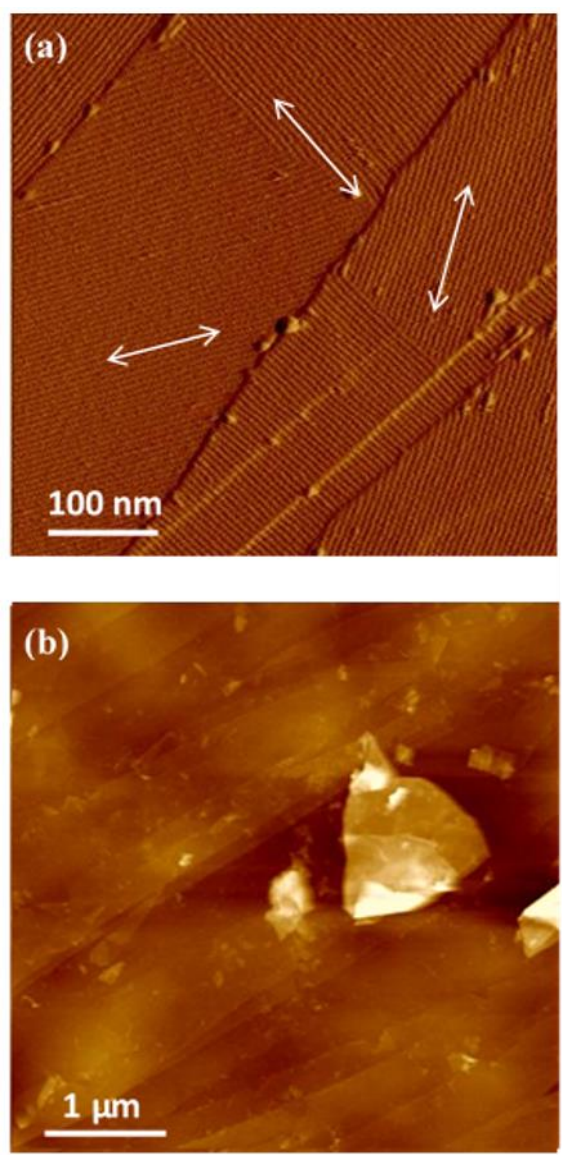

Figure 2. Tapping mode AFM images of HOPG: (a) after dropping with DMSO solution (amplitude image); (b) after sonication in DMSO solution (height image, Z-scale: 50 nm).

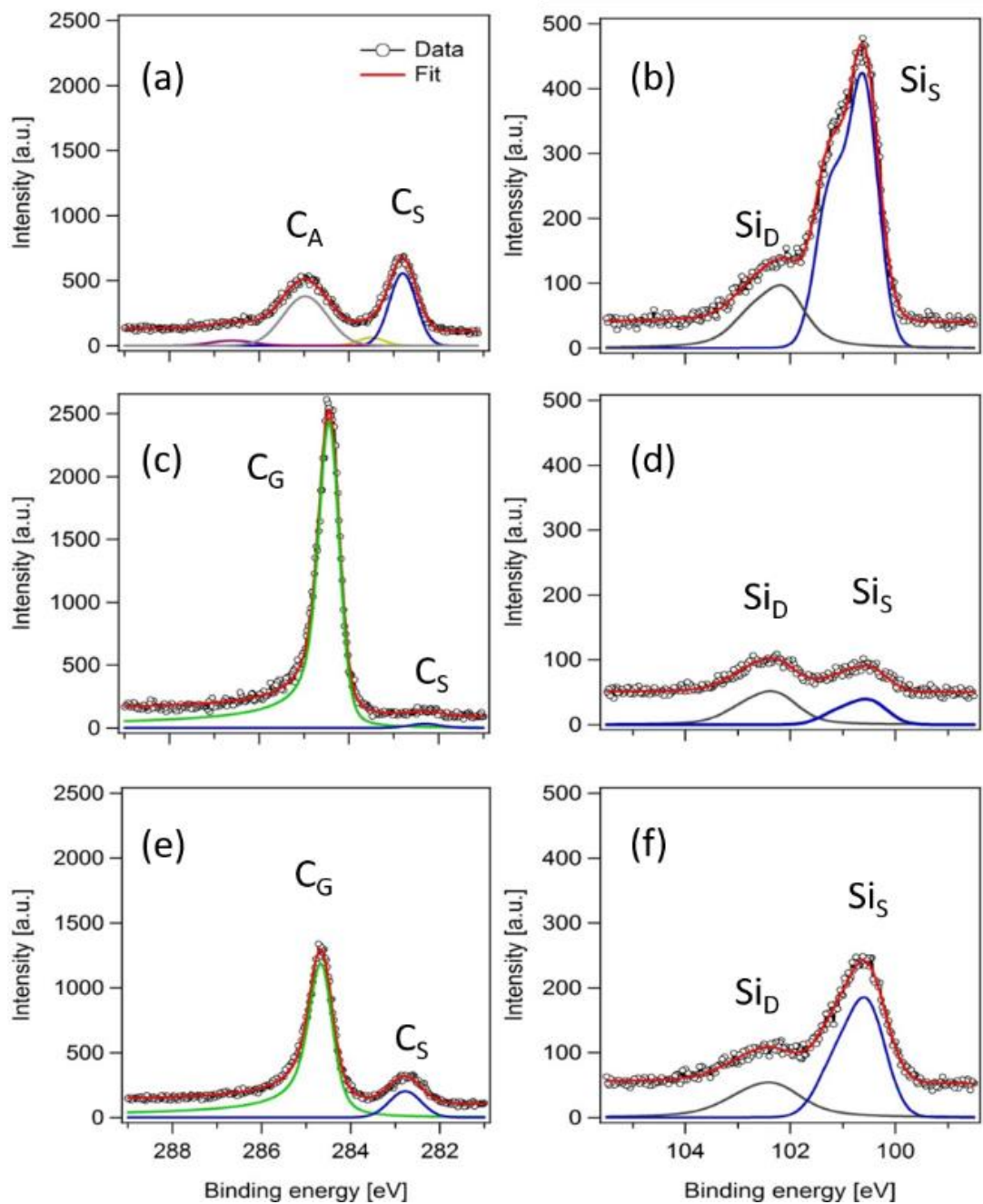

Figure. 3 XPS core level spectra. Left panels: C1s spectral region. Right panels: Si2p spectral region. (a-b) SiC substrate, (c-d) FLG, pristine, (e-f) FLG, after sonication in DMSO.

**XPS and Raman analysis**

Representative XPS results are reported in figure 3, which shows C1s and Si2p core level spectra for the pristine and treated FLG film. Spectra obtained on a bare SiC substrate are reported for comparison. The graphical choice adopted for the y-scales emphasizes the neat differences among the three cases. The figures show the decomposition in sub-components after a background (linear + Shirley-type) subtraction. Details on line-shape fitting are given below, for each panel.

Regarding the bare SiC, the C1s spectrum (Figure 3a) is characterized by two main sub-components which have been fitted by Voigt functions. The higher BE component $C_A$, located at 284.9±0.2 eV, was assigned to sp3 adventitious carbon and used as reference. The $C_S$ component, at 282.8±0.2 eV, was attributed to the SiC substrate after comparison with previous reports[52,53]. Two additional weak components at ~ 286.5 eV and 283.5 eV were included for a more accurate reproduction of the experimental profile. The former is perceptible also on raw data and can be likely assigned to contaminants. The latter, much weaker and correlated to the choice of background, is tentatively assigned to some SiC faults. The Si2p data (Figure 3b) also show two components. The 2p doublets have been modelled with Voigt functions, with a branching ratio of 1:2 and a spin orbit splitting of 0.6 eV. The doublet structure is fully evident only on the most intense $Si_S$ peak: the $2p_{3/2}$ sub-component is located at 100.6±0.1 eV BE and can be assigned to SiC[53,54]. The weaker component $Si_D$ is located at 102.1±0.2 eV BE. The severe broadening of $Si_D$ suggests the convolution of several contributions from defects. Regarding the pristine FLG, the C1s spectrum (figure 3c) is dominated by the intense, well-defined $C_G$ peak which exhibits the clear asymmetry toward higher binding energies typical of graphitic systems[55]. The peak was therefore fitted by a Doniach-Sunjic (DS) line shape function. The position, at 284.4±0.1 eV, is in agreement with previous works on FLG[52,56]. The FLG film is thick enough that the detection of photo-electrons belonging to the substrate is substantially suppressed: Only a faint $C_S$ component at ~ 282.5 eV is indeed reminiscent of SiC. Passing to the Si2p spectral region (Figure 3d) the $Si_S$ peak appears severely attenuated and its intensity is now lower than the $Si_D$ peak. It is also broadened, likely reflecting substantial alterations of the topmost layers of the substrate induced by the FLG formation process.

After sonication in DMSO (Figure 3e-f), the sharp decrease of the $C_G$ peak intensity on one side and the increase of the substrate-related peaks ($C_S$ and $Si_S$) on the other side testify the reduction of the FLG film thickness. The best fit positions were found at 284.6±0.1 and 282.8±0.2 eV for the $C_G$ and the $C_S$ components, respectively, which were reproduced by a Voigt profile. According to literature the small upward BE shift of the $C_G$ state is consistent with the thinning of the FLG film[52,53,57,58]. On the other hand, the $Si_D$ component appears much less affected by the sonication treatment. We may speculate that part of SiC faults could have hindered the formation of FLG.

We have applied well-known formulas[54,59,60] to determine the thickness of the FLG film based on the intensity of the film- and substrate-related XPS peaks, taking into account the photoelectron emission angle, attenuation lengths and elemental sensitivity factors. We exploited the intensity of FLG- and SiC-related peaks derived from C1s and Si2p spectra to obtain an estimate of the initial FLG film thickness and the reduction of thickness after sonication.

It turns out that 1h sonication in DMSO reduces the FLG thickness to ~40% of the initial thickness which can be estimated of the order of 3.5 nm (~ 12 layers), assuming a value for the attenuation length in FLG of ~ 2 nm for photoelectrons with kinetic energies of 1200-1300 eV[54].

Figure 4 shows the Raman spectra of FLG before and after 1-hour sonication in DMSO solution, together with data on bare SiC for comparison.

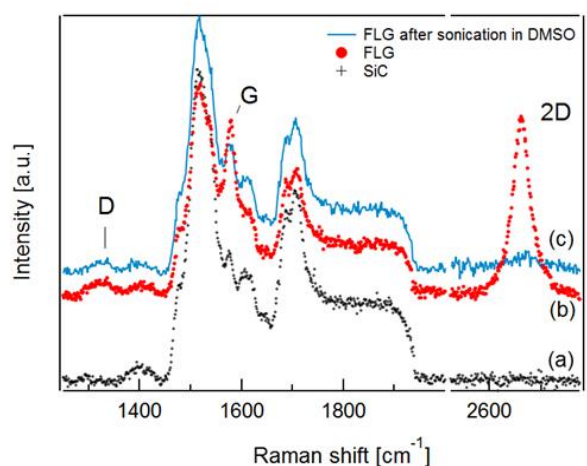

Figure 4 Representative Raman spectra. (a) SiC substrate (b) FLG, pristine (c) FLG, after sonication in DMSO solution. G, 2D and D peaks are characteristic of graphitic samples. The measurements were performed exploiting a 532 nm laser.

Regarding pristine FLG (curve b), a rather sharp 2D peak is observed at ~ 2665 cm$^{-1}$, out of the range of SiC overtones visible below 2000 cm$^{-1}$, in excellent agreement with literature[40,61,62]. The G peak is appreciable at ~1580 cm$^{-1}$, amid intense SiC contributions. The comparison with the SiC substrate spectrum (curve a) allows also to appreciate a low-intensity D peak at ~1340 cm$^{-1}$. The 2D peak does not show the multicomponent form typical of Bernal stacking of graphitic samples (e.g. HOPG)[40,63-65], and clearly resembles a single bell-shaped structure (38 cm$^{-1}$ f.w.h.m.), which is the fingerprint of the single-band dispersion relation of decoupled graphene layers as previously reported[40,66]. This finding can be attributed to the growth mechanism of FLG on SiC. Epitaxial layers obtained on the C-face of SiC contain rotational stacking faults which result in graphene sheets which are electronically decoupled, as it is the case to some extent of turbostratic graphite[52,67,68]. Therefore, the band structure of FLG films on the C-face on SiC is relatively similar to that of single layer graphene[40]. Sonication in DMSO results in a strong attenuation of G and 2D components (curve c). The decrease in intensity of the FLG peaks is accompanied by the increase in intensity of the SiC-related signals. Similar spectra were reported in literature for ultrathin films of epitaxial graphene with different thickness on SiC[54]. According to the analysis algorithm proposed in ref.[54], which is based on the evaluation of attenuation of SiC Raman peaks caused by the FLG film, we can estimate a thickness of the pristine FLG film of 12-13 layers, while the thickness reduces to about 5-6 layers after sonication. The Raman spectra of Figure 4 are therefore consistent with the XPS analysis, indicating an overall residual FLG thickness of a few layers after sonication.

Since for HOPG it is difficult to obtain direct information on exfoliation from XPS or Raman due to the persistent bulk graphitic signal, we resorted to the UV-Vis absorption spectroscopy analysis of the solution obtained after sonication of HOPG flakes. The absorption spectra (not shown) exhibited a maximum at ~ 270 nm, indicative of the π conjugated electronic structure of graphene sheets[14,65], thus confirming exfoliation also at the very low solution concentration adopted in the present work.

Regarding the "mild" treatment, XPS/Raman experiments didn't show significant changes in comparison to the pristine FLG film. However, in XPS mappings, a slight increase of the $Si_S/C_G$ ratio was observed in a few zones of the samples, eventually compatible with a local, very small reduction of the FLG film thickness. At variance with the sonication treatment, simple dropping was not able to induce significant exfoliation. Regarding molecular adsorption, wide-scan spectra occasionally showed the presence of trace signals of states related to the dropped molecules, like S (for both DMSO and L-methionine) or COOH states (for L-methionine) while the N1s state was never detected. Therefore, to the best of our analysis and under the UHV conditions needed for the XPS measurements, we can exclude the presence of organized, long range ordered molecular layers able to form ripples.

## Discussion

The ripple structure observed in the present study on FLG appear completely different in nature with respect to lamellar-type structures observed in several in-situ STM investigations at the solid-liquid interface between HOPG and neat liquid alkanes with both simple or substituted chains (e.g.[69-71]). In those studies the lamellar width was found to scale with the molecular chain length and the lamellae were ascribed to the formation of an ordered molecular layer at the solid-liquid interface.

Rippled domains were reported also in a STM study of the interface between HOPG and a methionine solution in octanol[72]. In that case the ripples were interpreted as rows of methionine dimers and the inter-row distance was found to increase with decreasing the methionine concentration.

We observed ripples in two situations: i) SP-type ripples on pristine FLG ii) LP-type ripples on FLG and HOPG surfaces rinsed and dried after "gentle" interaction with diluted DMSO or methionine aqueous solutions. The LP ripple structure was the same irrespective of the type of molecular solution and was the same that we observed in previous experiments on HOPG after gentle interaction with aqueous solutions of many molecules (proteins[26], ε-caprolactame[28], polyelectrolytes and L-methionine[27]). In some experiments on biomolecule interaction with HOPG we observed two exposed planes showing different ripple periodicity: LP-type ripples on the topmost terraces and SP-type ripples on the immediately lower level[26].

Considering these findings it is difficult to figure out the same adsorption pattern for molecules endowed with different structure, and molecular weights which span three orders of magnitude.

We have tried to get spectroscopic information on eventual adsorbed molecular layers by exploiting gentle spectroscopic ellipsometry (SE) measurements.

In favourable cases, through the use of difference spectra, SE was able to provide clear fingerprints of formation of molecular monolayers[46,73]. On substrates like gold[74] or $SiO_2$[75] we could detect the Soret band of sub-monolayer films of cytochrome c. We therefore performed SE measurements on HOPG after interaction with cytochrome c solution (data not shown). While we were able to detect well-defined ripples by AFM, we did not observe any clear spectral feature related to the molecular absorption. By applying the differential spectra analysis to the SE data we got tiny signals that could be compatible with both the adsorption of an ultrathin transparent layer or with surface roughening[27]. Further, the AFM observation of coexisting LP and SP rippled structures[26] points to exclude the assignment of ripples to molecular domains since quite peculiar phase segregated adsorption mechanisms should be invoked to account for coexisting domains.

The sum of our findings seems therefore to rule away quite definitely any assignment of the ripples to the sole organization of molecular material deposited from the solution.

In other experiments on the HOPG surface exposed to water, a surface morphology similar to the SP-type ripples has been observed[29-31]. The authors proposed an interpretation in terms of ordered domains of gas adsorbates ($N_2$ or $O_2$), which would eventually form after gas diffusion through the air-water interface and subsequent segregation at the hydrophobic-water interface as indicated by molecular dynamics simulations[76]. This interpretation was substantially endorsed in other works which studied exfoliated graphene on mica exposed to water[32] or epitaxial graphene on SiC, both mono- and bi-layer[33], and H-intercalated graphene on SiC[34], exposed to air. In the latter case[33,34], while the high density of disordered adsorbates prevented the observation of ripples on epitaxial graphene monolayer, the SP-rippled structure observed on bilayer and on H-intercalated graphene was ascribed to a gas layer diffusing to the surface through the adsorbed wetting layer. A recent paper[35] showed an interesting correlation between ripple orientation and friction anisotropy properties on exfoliated graphene. Further, the authors observed the SP-type ripples also on hBN and on graphene/hBN heterostructures; in the latter case a transition to LP-type ripples (from ~4 to ~6 nm periodicity) was observed upon sample thermal cycling at low temperature. The paper critically discussed the origin of the ripples starting from the adsorbed gas model[29]. The authors noted the lack of chemical analysis proving the nitrogen content of the stripes and pointed out that no explanation is given for the formation of nanometer-sized stripes instead of a homogeneous nitrogen layer. Within the adsorbate model, the authors[35] are in favour of the self-assembly of airborne and ubiquitous species which could account for the observation of the same stripe periodicity independently of the sample treatment. It is worth to mention that recent papers emphasized the influence of airborne organic contaminants on the wettability of graphitic surfaces[77-79].

The above mentioned studies[29,31,33-35] did not report spectroscopic evidence of the presence of adsorbed molecular layers.

We did not observe a $N_2$ layer as well in our XPS measurements. However, it should be noted that the UHV conditions adopted could have perturbed or even destroyed the delicate equilibrium conditions eventually necessary to the preservation of such a weakly bound surface gaseous layer.

We note that other recent AFM studies reporting on friction anisotropy domains on graphene and other atomically thin sheets on weakly adherent substrates[36-39] put forward a different interpretation, attributing rippled domains to out-of-plane deformations of ultrathin and weakly interacting films, i.e. almost 2D systems.

In particular, in ref.[39] a combined torsional AFM and ARPES study could demonstrate that ripples on exfoliated graphene on $SiO_2$ were aligned along the zig-zag direction of the hexagonal lattice; in the same study theoretical calculations indicated a ripple periodicity of 5.7 nm under a 10% compressive strain and a 0.16 meV/nm$^2$ lower energy for zig-zag directional ripples compared to armchair directional ones.

Therefore, while some works emphasize the role of airborne species diffusing through aqueous phase in the rippling process of graphitic surfaces[29-35], other works possibly point to other driving factors[36-39].

Interesting hints came from an experiment revealing that the deposition of triazine molecules from the vapour phase led to modify the stacking order in trilayer graphene from Bernal (ABA) to ABC stacking[25].

The different graphene stacking order in FLG and HOPG could be an interesting factor to be considered. During graphitization on the C-face of SiC, graphene layers experience some rotational freedom with respect to each other and lock in, on average, to preferred orientations. Indeed analysis of X-ray azimuthal scans indicated three preferred orientations of FLG relative to the SiC azimuth[68,80]. The low stacking order of epitaxial graphene results in an interlayer spacing inferred from X-ray reflectivity (0.3368 nm)[80] higher than that for crystalline graphite (0.335 nm)[67].

Graphene layers in pristine FLG can be therefore regarded as a sort of loosely interacting system, quasi-2D in character. In this respect, it is tempting to ascribe the SP rippled domains we observed on pristine FLG to a process of free energy minimization, similar to the surface rippling of free standing 2D materials[81-83].

In the case of pristine HOPG, the ordered Bernal-type stacking and the relatively higher van der Waals interactions stabilize the carbon planes in the planar configuration and inhibit the formation of ripples.

The interaction with aqueous solutions of organic compounds may modify the picture. We suggest that the relaxation from the SP to the LP-type ripples on FLG and the LP ripples observed on HOPG after "mild" treatment could be due the to the weakening of the van der Waals interplane interactions, induced e.g. by molecules bearing both apolar moieties, with high affinity for the carbon plane, and polar groups, highly interacting with the aqueous phase. Such a weakening could make the carbon planes even more similar to 2D systems possibly affecting the stacking and rotational order.

We note that the observation of LP and SP ripples on topmost and second level terraces points to the possible role of stacking and interplanar interactions while seems less encouraging regarding an interpretation in terms of layers formed by "adventitious" molecules.

When additional energy is transferred to the system through sonication, the reduction of van der Waals interactions proceeds further and, even though very diluted solutions are used, graphene planes are peeled off, in a similar way as previously reported in literature for liquid exfoliation of graphene in proper organic solvents/solutions. Interestingly when exfoliation occurs, rippled domains are no more observed on FLG, likely due to the higher defect density resulting from sonication.

Several reports investigated the mechanical response of loosely bound quasi-2D systems following STM or AFM tip perturbation[84-87] and reported local distorsion of the Moiré pattern observed on supported graphene. At the computational level, Duan et al.[88] developed a continuum model of a graphene sheet which accounts for ripple formation under the influence of a shear stress; in that case, however, ripple periodicity was found to be modulated by shear stress intensity. In Friction Force Microscopy studies on HOPG, Rastei et al ascribed the observation of rippled domains to a tip-induced puckering process[89,90].

Further, the AFM operation mode has been reported to influence the observation of ripples. Giessibl and coworkers[33,34] observed ripples on bilayer graphene and hydrogen-intercalated graphene on SiC with AFM operated in force-modulation mode (FM-AFM). Hwang and coworkers observed ripples in PeakForce- and FM-AFM, but not in TM-AFM[29-32]. In the present study as well as in ref[35] ripples could be resolved in TM-AFM. The capability of observing ripples therefore seems to depend on subtle details of the tip/surface interaction conditions.

## Conclusions

From the discussion of our results and considering data reported in literature it emerges that at least three, possibly interrelated, aspects can be into play for the observation of ripples on graphitic substrates: (i) the interplane order and stacking (loosely interacting, quasi 2D systems could chose rippling for energy minimization), (ii) a soft interaction between surface and aqueous phase, either solution of bio-organic molecules or water with dissolved gaseous species, (iii) the perturbation of the graphene surface by the scanning AFM tip.

We would exclude the assignment of ripples to organic molecules that adsorb on the surface from the solution forming ordered domains, as no experiment was able to gather spectroscopic evidence of the presence of an extended molecular layer on the surface.

We do not exclude the interpretation of ripples in terms of corrugation of a molecular layer formed by gaseous "contaminants" coming from the ambient or dissolved in water. The research work by Hwang and coworkers[29-32], exploring the dynamics of ripple formation, supports this interpretation. Nevertheless, as already mentioned, the simultaneous observation of SP and LP ripples on adjacent planes[26] of HOPG seems hard to be rationalised in terms of the sole molecular organization.

As mentioned above SP ripples have been observed in 2D materials different from graphene and have been ascribed to environmental adsorbates which would organize similarly on very different substrates[35].

However a different interpretation can be attempted. The liquid phase interacting with the topmost surface layers can weaken the carbon plane coupling or even to some extent intercalate between carbon planes.

Ripples could therefore be regarded not as the direct image of the molecular layer but eventually as the result of a free-energy minimization process of loosely interacting quasi 2D graphene layers, promoted by molecules which could either weaken the interplane interaction and/or act as surfactants/intercalants. In this respect it is worth to mention that noble gas intercalation between graphene and Ir[91] or Pt[92] surfaces has been reported to produce graphene deformations at the nanoscale while water intercalation has recently been reported between graphene and a hydrophilic substrate[93].

To get further insight into the origin of ripples, it could be worth to look for rippling processes on other 2D materials to evaluate how lattice constants, interplane interactions and stacking order can eventually modulate rippling.

Definite conclusions about the actual presence and role of weakly bound adventitious species needs a convincing spectral characterization, which appears difficult to obtain for this kind of systems.

Finally we note that, though the AFM operational conditions to observe ripples could be critical, the nanopatterns, once formed, can be even exploited as templates to guide the oriented deposition of highly anisotropic objects, such as amyloid-like fibrils obtained by nanoparticle-induced protein aggregation[94].

## Acknowledgements

Financial support from the Ministero dell'Istruzione, Università e Ricerca (Project no. PRIN 20105ZZTSE_003) is acknowledged. Part of the research leading to these results has received funding from the European Union Seventh Framework Program under grant agreement no. 604391 Graphene Flagship.

## Conflicts of interest

There are no conflicts to declare.